\documentstyle[aps,graphicx,epsf,amssymb,eqsecnum,pre,multicol]{revtex}
\title{Thixotropy in  macroscopic suspensions  of spheres}
\author{C. V\"oltz, M. Nitschke, L. Heymann$^*$ and I. Rehberg}

\address{Experimentalphysik V, 
Universit\"at Bayreuth, D-95440 Bayreuth, Germany}

\address{$^*$ Technische Mechanik und Str\"omungsmechanik, Universit\"at Bayreuth, D-95440 Bayreuth, Germany }

\date{\today}
\begin{document}
\maketitle
\begin{abstract}
An experimental study of the viscosity  of a  {\it macroscopic} suspension, i.e. a suspension for which Brownian motion can be neglected, under steady shear is presented. The suspension is prepared with a high packing fraction and is  density-matched in a  Newtonian carrier fluid.  The  viscosity of the suspension  depends on the shear rate and the time of shearing.   It  is shown for the first time that a {\it macroscopic} suspension shows thixotropic viscosity, i.e.\ shear-thinning with a long relaxation time as a unique function of shear. 
The relaxation times show a systematic decrease with increasing shear rate.  These relaxation times are larger when decreasing the  shear rates, compared to those observed after increasing the shear. The time scales involved are about 10000 times larger than the viscous time scale $\tau_{visc}=a^2/\nu$ and about 1000 times smaller than the thermodynamic time scale $\tau_{therm}=Pe/\dot{\gamma}$.  
The  structure of the suspension at the outer cylinder of a viscometer is monitored with a camera, showing the formation of a hexagonal structure.  The temporal decrease of the viscosity under shear  coincides with the  formation of this hexagonal pattern. 
\\
\\
PACS number(s): 83.60.Fg, 83.60.Rs, 82.70.Kj, 47.27.Pa, 47.50.+d, 83.50.Ax
\end{abstract}

\begin{multicols}{2}

\section{Introduction}
\label{sec:1}
The rheology of suspensions has been studied for many years, mainly because of its obvious importance in a wide range of industrial applications (see for example \cite{barnes}, \cite{ungarish}). 

A lot of work has been done especially for {\it colloidal} suspensions, i.e.\ suspensions where Brownian motion cannot be neglected \cite{everett}, \cite{barnes-hutton}, \cite{nommensen}, \cite{ackerson1988}, \cite{ackerson1990},  \cite{tomita}, \cite{gillespie}, \cite{krieger},  \cite{wagstaff}, \cite{dekruif}, \cite{willey}, \cite{hoffman}, \cite{woodcock}, \cite{vanderwerff}. 
In these {\it colloidal} suspensions shear-thinning can be observed, i.e.\ the viscosity of the fluid decreases with increasing shear rate \cite{schowalter}, \cite{gupta}, \cite{kissa}, as well as shear-thickening  for higher shear rates. It is known, that {\it colloidal} suspensions also show thixotropic behavior \cite{vannostrand}, \cite{barnes2}, i.e.\  the continuous reduction in viscosity with time of shearing and the subsequent recovery under reduced shear  \cite{barnes-hutton}, \cite{gupta},  \cite{barnes2},  \cite{macosko}, \cite{tscheuschner}.
The idea is proposed, that the change in viscosity is caused by the formation of structures within a {\it colloidal} suspension under shear  \cite{ackerson1988}, \cite{ackerson1990}, \cite{barnes2}, \cite{macosko}. 
It  is assumed that for {\it colloidal} suspensions the dynamics is dominated by Brownian diffusion \cite{willey}, \cite{barnes2},  \cite{verberg}. 

A measure for the relative importance of  shear forces compared to Brownian motion is given by the P\'eclet-number $Pe= 6\pi\eta\dot{\gamma}r^3/(k_B T)$,  where $r$ is the radius of the particles. 
Whereas {\it colloidal} suspensions have been investigated intensively up to a P\'eclet-number of 100, few experimental data exist for {\it macroscopic} suspensions, i.e.\ for suspensions of higher P\'eclet-number where Brownian motion can be neglected.
It is assumed  that for high P\'eclet numbers  the shear-thinning region shifts rapidly to extremely low values and essentially becomes experimentally inaccessible \cite{macosko}, \cite{ferguson}. The lack of Brownian motion for these large particles is believed to eliminate the reversibility in structural changes and should complicate experiments \cite{barnes2}, \cite{macosko}. These difficulties explain  the dearth of experimental data on the behavior of {\it macroscopic}, i.e.\ non-Brownian suspensions.

For this class  interesting effects have been observed  for {\it non-Newtonian} carrier fluids (see e.g. \cite{lyon}, \cite{giesekus}, \cite{michele}, \cite{petit}).
Suspensions in {\it Newtonian} media appear to have been almost neglected. Some works showed the difficulties arising when trying to investigate the behavior of concentrated suspensions. It was demonstrated that shearing the macroscopic suspensions in a concentric cylinder-viscometer of the Searle-type leads to shear migration effects, i.e.\ the particles migrate in nonhomogeneous shear flows from high shear rates to low shear rates \cite{gadala}, \cite{leighton}, \cite{phillips} \cite{massoudi}. For the geometry used in those experiments it  meant that the particles even migrated out of the sheared volume and the particle concentration decreased. Since for high packing fractions the viscosity depends  strongly on the concentration, this reduction in concentration was measured. So from these measurements no estimate of the real viscosity  of the suspension could be made. 

In another experiment \cite{gondret} a macroscopic suspension was  exposed to oscillatory shear and the decrease of the viscosity with time was measured. In this experiment there was  a hint for an increase in the organization of the particles with shearing time. Nevertheless the suspension was not density matched, so it is not proved that the decrease in viscosity is not, at least partially, due to sedimentation.

The present paper presents experimental results on the behavior of {\it macroscopic} suspensions under steady shear at $\dot{\gamma}=50/$s of a P\'eclet-number $Pe \approx 2.1\cdot 10^9$ when using the viscosity of the suspension or  $Pe \approx 8.3\cdot 10^7$ when using the viscosity of the carrier fluid,  i.e.\ the P\'eclet number is seven orders of magnitude larger than the P\'eclet numbers which are covered by the experiments  with colloidal suspensions. We find evidence for a thixotropic behavior of the suspension with extremely large relaxation times. The relaxation times are about 10000 times larger than the viscous time scale $\tau_{visc}=a^2/\nu$ and about 1000 times smaller than expected from the P\'eclet number $\tau_{therm}=Pe/\dot{\gamma}$. This makes it unlikely that a hydrodynamic or thermodynamic interaction is responsible for the effect. One can thus speculate that the time scale is dominated by ordering effects within the densely packed granular network provided by the beads \cite{veje}. 
The decrease of the viscosity under shear  coincides with an  increase in order of the particles.

The paper is organized as follows: In the following section the experimental setup is described, in Section \ref{results} the experimental results are discussed. A final section contains the conclusions.

\section{Experimental setup}
\label{setup}

\begin{figure}[h]
\noindent
\begin{minipage}{8.6cm}
\epsfxsize=8.6cm
\epsfbox{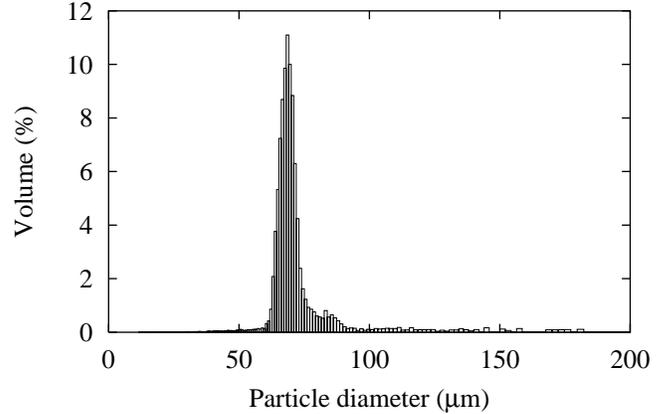}
\caption{Size distribution of the used particles, measured from conductivity measurements.}
\label{fig1}
\end{minipage}
\end{figure}

The suspension investigated consisted of nearly monodisperse glass beads
(W\"urth Ballotini MGL) dispersed in an aqueous solution of zinc iodide $ZnI_2$.
The glass beads are spherical and consist mainly of $SiO_2$ and have a material density of $\rho=2.45$ g/cm$^{3}$.
The particles are sieved in order to obtain a well defined particle size distribution which is then measured with  a Multisizer 2 (Coulter). The particle size distribution (PSD) is shown in  Fig.~\ref{fig1}. This measurement of the PSD is based on the principle of monitoring the electrical current between two electrodes immersed in a conductive liquid on either side of a small aperture, through which the suspension of particles is forced to flow. As a particle passes through the aperture, it changes the impedance between the electrodes and produces an electrical pulse of short duration with an amplitude essentially proportional to the particle volume. The arithmetic mean  of the particle size  is 61.13 $\mu$m and the standard deviation is 18.71 $\mu$m. 
An alternative measurement of the particle size distribution was  based on laser light diffraction experiments. These measurements confirm the results in Fig.~\ref{fig1}.
Before the glass beads were suspended in the solution they were  washed 8 times in totally demineralized water, in order to purify them from any residues and dried afterwards.

To stabilize the suspension, i.e.\ to avoid sedimentation during the time-consuming rheological experiments, the density of the continuous phase has been matched to that of the glass beads. 
Hence the particles were suspended in almost saturated  $ZnI_2$-solution, which had the same density $\rho=2.45$ g/cm$^{3}$ at $T=20.0^{\circ}$C as the glass beads. (The maximal density of the solution is about 2.6 g/cm$^{3}$.)  The density-matched, i.e. stabilized, suspension was prepared with the packing fraction $\phi=50.49$\%.
It was checked at $T=20.0^{\circ}$C for several hours that the  suspension stayed homogeneous and the glass beads did not sediment to the bottom or rise to the top of the vessel. 
The carrier fluid, the pure $ZnI_2$-solution, is a Newtonian solution with a   dynamic viscosity of  12 mPas at 20.0$^{\circ}$C.

The rheological measurements were performed with a rheometer SR-5000 (Rheometric Scientific, Inc.) with a rotating inner cylinder  in a fixed outer cylinder.  The inner cylinder has  a cone at the lower end.  Due to the relatively large particles a concentric cylinder geometry  with a ratio $\delta=R_i/R_o= 0.922$ of the inner cylinder radius $R_i=14.75$~mm   to the outer cylinder radius $R_o=16.0$~mm  and a gap width of $a$=1.25 mm has been used. The  length of the inner cylinder is 44.3~mm.  The shear stress in the small gap is constant within 15\% for a Newtonian fluid.  The thixotropic behavior was  investigated in the controlled shear rate mode by a stepwise increase and subsequently stepwise decrease of the shear rate $\dot{\gamma}$  and registration of the shear stress $\sigma$ and the apparent viscosity $\eta_{app}=\sigma/\dot{\gamma}$. The temperature during the whole  measurements was stabilized to 20.0 $^{\circ}$C.

The shear rates are below  $\dot{\gamma} = 500/s$, which corresponds to a Taylor number \cite{gyon}
$Ta =\Omega^2 R a^3\nu^{-2}\approx 11.3$,
where $\Omega$ is the angular velocity and $R=(R_i+R_o)/2$.    Hence the experiments are performed far away  from the onset of Taylor-vortices at $Ta$=1712. 

%--------------------------------------------------------------------------------------

In order to allow for an optical investigation of the structure-formation at the outer cylinder of the suspension, a shearing-apparatus was built which had the same gap width $a$=1.25 mm and almost the same ratio $\delta$=0.916  as the rheometer. The outer cylinder had a radius of 15.0 mm and was made out of glass to allow the view into the suspension layer at the outer cylinder. The  inner cylinder did not have a conic but a flat lower end. Images were taken during shearing. In addition the torque of the motor was monitored, which is  a measure for the viscosity of the suspension.

\section{Experimental results}
\label{results}
\begin{figure}[h]
\noindent
\begin{minipage}{8.6cm}
\epsfxsize=8.6cm
\epsfbox{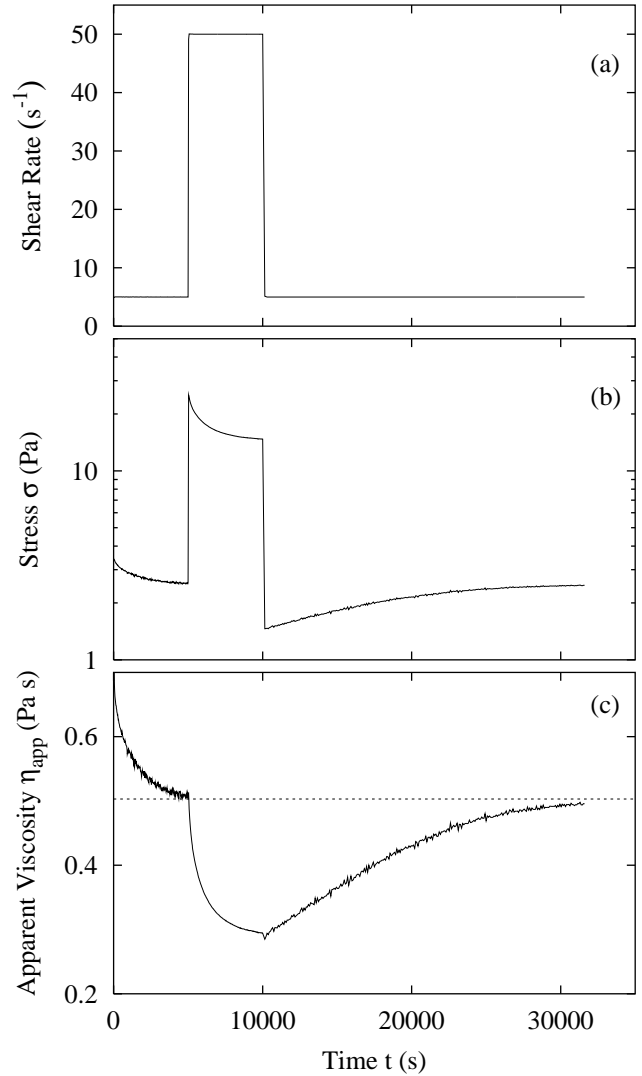}
\caption{Thixotropic behavior of a suspension of packing fraction of 50.49\%. (a) Driven shear rate ramp, (b) stress, (c) dynamic viscosity of the suspension}
\label{fig2}
\end{minipage}
\end{figure}

The  following sequence of shear rate steps  was  used: $\dot{\gamma}=$ 0/s, 5/s, 50/s, 5/s, 0/s. The resulting shear stresses  and the apparent viscosities $\eta_{app}$ were measured  as a function of time. Fig.~\ref{fig2}(a) shows the  shear rate steps together with the system response, i.e.\ (b) the stress and (c) the dynamic viscosity $\eta$. When $\dot{\gamma}$ is set from 0 to 5/s an exponential decay of the viscosity to the final viscosity is observed. Then the shear rate is set to 50/s and another exponential decay to another apparent viscosity is observed. This final value of the viscosity is smaller than the final value at $\dot{\gamma}=$5/s.   After setting $\dot{\gamma}$ back to 5/s the former final viscosity at this shear rate is gradually reached again. This is the behavior of a thixotropic material.

From this measurement the relaxation times $\tau$ are obtained from a fit to 

\begin{equation}\label{relaxation-times}
\eta(t)= a \cdot exp(-t/\tau) + \eta_{final}  
\end{equation}

\begin{figure}[h]
\noindent
\begin{minipage}{8.6cm}
\epsfxsize=8.6cm
\epsfbox{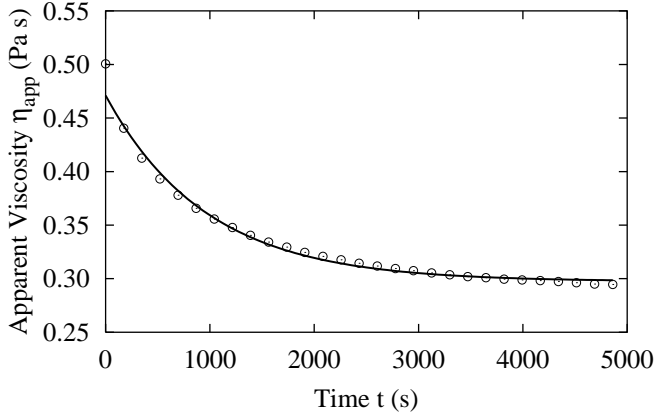}
\caption{Exponential fit to the decay of the dynamic viscosity with time for the step from $\dot{\gamma}=$5/s to $\dot{\gamma}=$50/s. For the reason of clearness only every 10th data point is shown.}
\label{fig3}
\end{minipage}
\end{figure}

\begin{figure}[h]
\noindent
\begin{minipage}{8.6cm}
\epsfxsize=8.6cm
\epsfbox{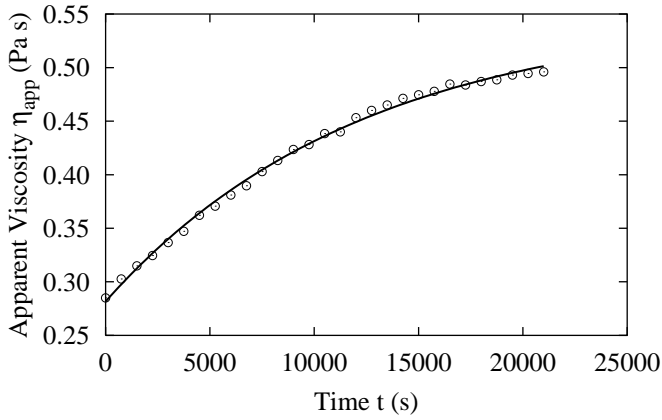}
\caption{Exponential fit to the growth of the dynamic viscosity with time for the step from $\dot{\gamma}=$50/s to $\dot{\gamma}=$5/s. For the reason of clearness only every 10th data point is shown.}
\label{fig4}
\end{minipage}
\end{figure}

Fig.~\ref{fig3} and Fig.~\ref{fig4} show the fits for $\dot{\gamma}=$50/s (from $\dot{\gamma}=$5/s) and for $\dot{\gamma}=$5/s (from $\dot{\gamma}=$50/s).

For the first shear rate $\dot{\gamma}=$5/s the relaxation time is $\tau$=1276 s.
For the shear rate $\dot{\gamma}=$50/s the relaxation time is $\tau$=968 s and for the gradual rise in viscosity for $\dot{\gamma}$=5/s the relaxation time $\tau$=12182 s, i.e.\ it is one order of magnitude larger than for the decay in viscosity at the same shear rate.

The ramp for these shear rates was not applied in a repeated sequence with the same suspension, but was applied four times for freshly prepared suspensions  and the thixotropic behavior of the suspensions was reproducible.  For the following shear rates the ramp  was applied once with alternating jumps from higher to lower and lower to higher shear rates.

\begin{figure}[h]
\noindent
\begin{minipage}{8.6cm}
\epsfxsize=8.6cm
\epsfbox{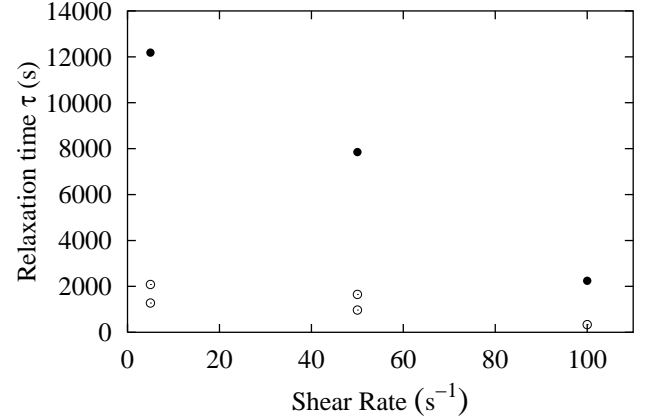}
\caption{Dependence of the relaxation time $\tau$ on the shear rate. Filled symbols represent steps from higher shear rates to lower shear rates, open symbols represent steps from lower shear rates to higher shear rates.}
\label{fig5}
\end{minipage}
\end{figure}

This evaluation of the relaxation times was also done for other measurements for both possible cases: coming to the appropriate shear rate from a lower shear rate (step from $\dot{\gamma}$=0 to 5/s, from 5/s to 50/s and from 10/s to 100/s) or coming from a higher shear rate (step from $\dot{\gamma}$=100/s to 50/s and 500/s to 100/s). The duration of shearing was 3 hours when coming from a lower shear rate and 6 hours when coming from a higher shear rate, respectively. 
The relaxation times show   a systematic decrease for increasing shear rates.  For decreasing steps from higher shear rates to lower shear rates the relaxation times are larger than for increasing steps. 
All results are shown in Fig.~\ref{fig5}. 

%------------------------------------------------------------------------------------
%------------------------------------------------------------------------------------

\begin{figure}[h]
\noindent
\begin{minipage}{8.6cm}
\epsfxsize=8.6cm
\epsfbox{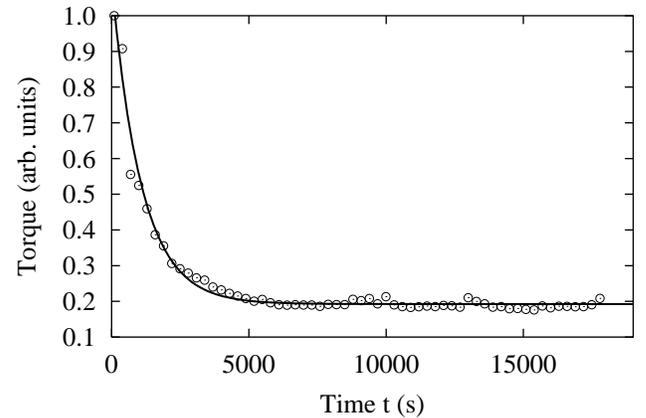}
\caption{Decrease of the torque of the motor, which is a measure of the viscosity of the suspension under shear. The torque is averaged over 100 s. Every  3rd  data point is shown.}
\label{fig6}
\end{minipage}
\end{figure}

Using the optical setup, the torque of the motor was monitored. Fig.~\ref{fig6} shows the decrease of the torque averaged over 100 s. An exponential fit according to  equ.~\ref{relaxation-times} gives the relaxation time of 1089 s, which is close to the one measured with the viscometer. Although the signal to noise ratio is worse than with the viscometer this allows the interpretation that the same mechanism is responsible for the torque reduction in both apparatuses. 

\begin{figure}
\noindent
\begin{minipage}{17.2cm}
\epsfxsize=17.2cm
\epsfbox{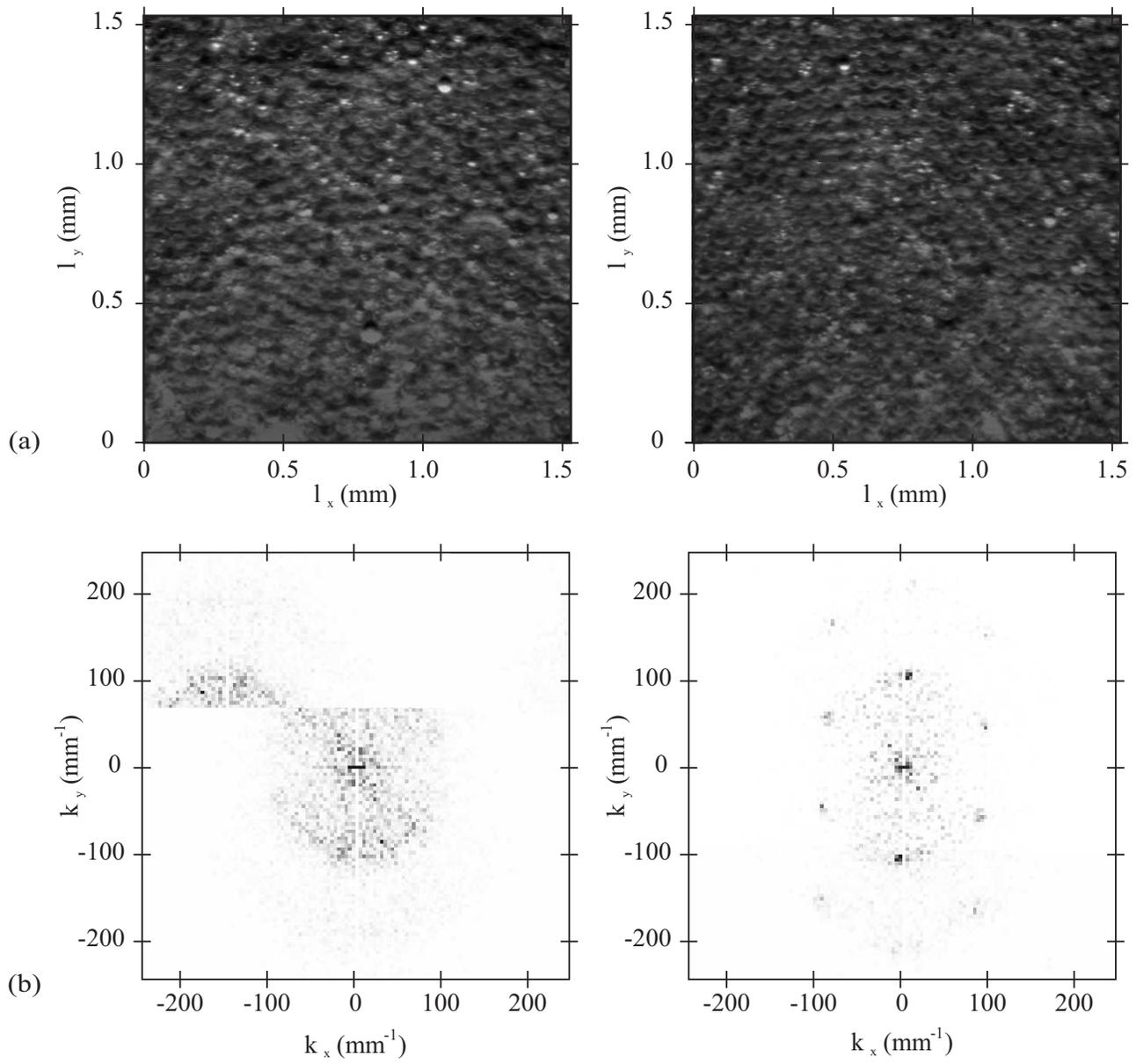}
\caption{(a) Images of the suspension at the outer cylinder at $t=0$ and $t=62$ min.  
(b) 2D-Fourier spectra of the images. At $t=0$ no structure is exhibited, at $t=62$ min a hexagonal structure is dominant. The shear is applied in the $x$-direction and the axis of the rotating cylinder is parallel to $y$.}
\label{fig7}
\end{minipage}
\end{figure}

Images with a Long Range Microscope (Questar QM100) which is connected to a CCD-camera  using an exposure time of about 1 ms  are taken from the suspension at the outer cylinder every 10 s during shearing at $\dot{\gamma}=$50/s.  The starting time of the shearing is defined by $t=0$. The images  show a cantle of 1.53 x 1.53 mm and are enhanced in contrast. Fig.~\ref{fig7}(a) shows images of the suspension at the outer cylinder at $t=0$ and at $t=62$ min. There is almost no difference observable in real space but the 2D-Fourier spectra of the images (Fig.~\ref{fig7}(b)) show a hexagonal structure after 62 min of shearing.

\onecolumn
\twocolumn

\begin{figure}[h]
\noindent
\begin{minipage}{8.6cm}
\epsfxsize=8.6cm
\epsfbox{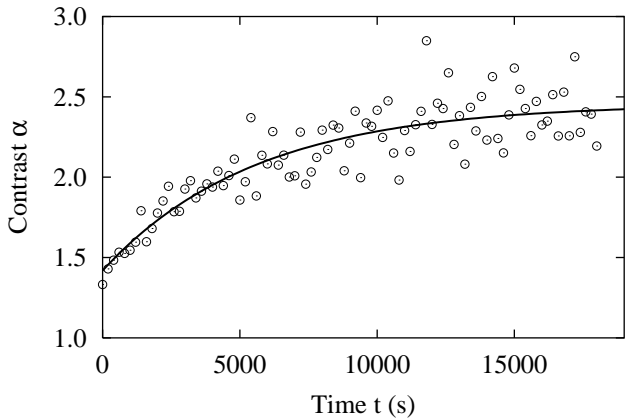}
\caption{Increase of the contrast $\alpha$, which is a measure of the order within the suspension. The line is a guide to the eye.}
\label{fig8}
\end{minipage}
\end{figure}

The increase of this hexagonal structure is quantified by calculating  a measure for the contrast  $\alpha$ as the root-mean-square normalized by the mean intensity of the spectral intensities for wave numbers $k=\sqrt{k_x^2+k_y^2}$ within a range of   $94/$mm $\leq k \leq 106/$mm, i.e.
\begin{equation}\label{alpha}
\alpha = \frac{1}{\overline{i}} \cdot \sqrt{\frac{\sum(\overline{i}-i(k))^2}{n}},
\end{equation}
where $i$ is the intensity in the spectrum and $n$ is the number of data points.
This  contrast is  averaged over 20 images in order to reduce  the noise. The unstructured spectra have a lower contrast  $\alpha$ than the structured spectra, i.e.\  $\alpha$  is a  measure for the structure of the suspension at the outer wall of the cylinder. Fig.~\ref{fig8} shows the temporal increase of that number.

%-----------------------------------------------------------------------------
\section{Conclusion}
\label{conclusions}
An experimental study of the viscosity  of a  {\it macroscopic} suspension of high packing fraction which is {\it density-matched}  in a {\it Newtonian} carrier fluid  under {\it steady} shear is  presented. It is demonstrated   that  {\it macroscopic} suspensions can be considered as thixotropic materials.

Furthermore  relaxation times at the order of hours were observed. They show   a systematic decrease for increasing shear rates.  For decreasing steps from higher shear rates to lower shear rates the relaxation times are larger than for increasing steps. 

During the decrease of the viscosity under shear a structure-formation was observed. The Fourier spectra show that the system forms a hexagonal structure within time. Both effects, i.e.\ the decrease of viscosity and the increase of patterns, coincide. 

We cannot provide an explanation for the physical origin of the thixotropic behavior. We note that the relaxation times are about 10000 times larger than the viscous time scale $\tau_{visc}=a^2/\nu$ and about 1000 times smaller than the thermodynamic time scale $\tau_{therm}=Pe/\dot{\gamma}$. This makes it unlikely that  hydrodynamic or thermodynamic interactions are responsible for the effect. 

One can thus speculate that the time scale is dominated by ordering effects within the densely packed granular network provided by the beads \cite{veje}.
In particular,  we believe that the friction between the wetted surfaces of spheres is primarily responsible for the measured effective viscosity. In a densely packed granulate material  that friction will be dominated by the fact   that individual particles are subject to frustration if they feel more than one particle contact, which tends to increase the effective viscosity. Ordering under shear might mean that the particles reduce the mutual friction by reducing the number of frustrated contacts which in turn might cause a lower viscosity. This idea is related to  the concept of dilatancy,  which is fundamental in the description of fluid properties of granular matter. Additional investigations using three-dimensional imaging techniques and theoretical calculations are clearly necessary to substantiate these speculations.

\section*{Acknowledgements}
We like to thank Nuri Aksel for helpful discussions and Markus Horn for his help during the measurements. 
The experiment was supported by Deutsche Forschungsgemeinschaft through Re588/12.

\end{multicols}
\end{document}